\documentclass[aps,prd,nofootinbib,superscriptaddress,preprint]{revtex4}
\usepackage{amssymb}
\usepackage{amsmath,color}
\usepackage{epsfig}
\usepackage{graphicx,epsf,epsfig}
\usepackage{bbm}
\usepackage{subfigure}
\usepackage{multirow}
\usepackage{setspace}
\usepackage{verbatim}
\usepackage{mciteplus}

\newcommand{\ba}{\begin{array}{c}}
\newcommand{\baz}{\begin{array}{cc}}
\newcommand{\bad}{\begin{array}{ccc}}
\newcommand{\bav}{\begin{array}{cccc}}
\newcommand{\baf}{\begin{array}{ccccc}}
\newcommand{\ea}{\end{array}}

\def\be{\begin{equation}}
\def\ee{\end{equation}}
\def\gs{\mathrel{
   \rlap{\raise 0.511ex \hbox{$>$}}{\lower 0.511ex \hbox{$\sim$}}}}
\def\ls{\mathrel{
   \rlap{\raise 0.511ex \hbox{$<$}}{\lower 0.511ex \hbox{$\sim$}}}}
\renewcommand{\baselinestretch}{1.2}
\newcommand{\bea}{\begin{equation} \begin{array}{c}}
\newcommand{\eea}{ \end{array} \end{equation}}

\def\slc#1{\setbox0=\hbox{$#1$}           
    \dimen0=\wd0                                 
    \setbox1=\hbox{/} \dimen1=\wd1               
    \ifdim\dimen0>\dimen1                        
       \rlap{\hbox to \dimen0{\hfil/\hfil}}      
       #1                                        
    \else                                        
       \rlap{\hbox to \dimen1{\hfil$#1$\hfil}}   
       /                                         
    \fi}
\begin{document}
 \mbox{}\vspace{1cm}
\title{Impact of massive neutrinos on the Higgs self-coupling and
electroweak vacuum stability}

\author{Werner Rodejohann}
\email{werner.rodejohann@mpi-hd.mpg.de}

\author{He Zhang}
\email{he.zhang@mpi-hd.mpg.de}

\affiliation{Max-Planck-Institut f{\"u}r Kernphysik, Saupfercheckweg
1, 69117 Heidelberg, Germany \vspace{1cm}\\[1cm] } \mbox{ } \\[1cm]

\begin{abstract}
\noindent The presence of right-handed neutrinos in the type I
seesaw mechanism may lead to significant corrections to the RG
evolution of the Higgs self-coupling. Compared to the Standard Model
case, the Higgs mass window can become narrower, and the cutoff
scale become lower. Naively, these effects decrease with decreasing
right-handed neutrino mass. However, we point out that the unknown
Dirac Yukawa matrix may impact the vacuum stability constraints even
in the low scale seesaw case not far away from the electroweak
scale, hence much below the canonical seesaw scale of $10^{15}$ GeV.
This includes situations in which production of right-handed
neutrinos at colliders is possible. We illustrate this within a
particular parametrization of the Dirac Yukawas and with explicit
low scale seesaw models. We also note the effect of massive
neutrinos on the top quark Yukawa coupling, whose high energy value
can be increased with respect to the Standard Model case.
\end{abstract}

\maketitle


\section{Introduction}

Direct observational constraints on the Higgs boson
\cite{Aad:2012si,Chatrchyan:2012tx,Tevatron:2012} suggest that, if
it exists, its detection should be close. Apart from direct limits,
indirect limits from electroweak precision data have been obtained
\cite{bla:2010vi}. Another approach to study Higgs properties, on
which we focus here, analyzes the high energy values of the Higgs
self-coupling
\cite{Linde:1975sw,*Linde:1975gx,*Weinberg:1976pe,*Politzer:1978ic,*Hung:1979dn,*Cabibbo:1979ay,*Flores:1982rv,*Lindner:1985uk,Sher:1988mj,Lindner:1988ww,*Casas:1994qy}.

Within the framework of the Standard Model, the one-loop
renormalization group equation (RGE) of the Higgs self-coupling
$\lambda$ is given by $(4\pi)^2 \frac{{\rm d}\lambda}{{\rm d}\ln
\mu} = \beta_\lambda$, with~\cite{Gunion:1989we,*Dawson:1998yi} \bea
\label{eq:beta}
\beta_\lambda  =  6 \lambda^2 - \lambda\left( 3 g^2_1 +9 g^2_2 \right) + \left( \frac{3}{2}g^4_1 + 3 g^2_1 g^2_2 + \frac{9}{2}g^4_2 \right)  \\
+ 4 \lambda {\rm tr} \left(3 Y^\dagger_u Y_u + 3 Y^\dagger_d Y_d +
Y^\dagger_\ell Y_\ell \right) - 8 {\rm tr} \left[ 3 \left(
Y^\dagger_u Y_u \right)^2 + 3 \left( Y^\dagger_d Y_d \right)^2+
\left( Y^\dagger_\ell Y_\ell \right)^2 \right]  . \eea Here $g_i$
are the gauge couplings and $Y_f$ (for $f=u,d,\ell$) denote Yukawa
coupling matrices of up-quarks, down-quarks and charged leptons. Two
important limits exist when one runs $\lambda$ from the weak scale
to larger scales:

\begin{itemize}
\item in the limit of small $\lambda$, the top quark Yukawa drives
$\lambda$ to negative values, and the Higgs potential is no longer
bounded from below.  This leads to the vacuum stability bound;

\item in the limit of large $\lambda$, the Higgs self-coupling drives
$\lambda$ to even larger values, and the presence of a
non-perturbative coupling leads to the triviality bound.
\end{itemize}
These two aspects imply lower and upper values of the Higgs mass,
depending on the embedding (or cutoff) scale $\Lambda$ of the
theory. In particular, for interesting values of Higgs mass around
$125 \pm 2$ GeV, latest numerical analyses in the literature
\cite{Holthausen:2011aa,EliasMiro:2011aa,Xing:2011aa} indicate that
$\Lambda$ should be around $10^9 \sim 10^{11} ~{\rm GeV}$ including
uncertainties from the top quark mass, the strong coupling and
higher order effects. Slightly larger Higgs masses
would cause no conceptual problems up to the Planck scale,
even larger values would lead to
non-perturbative $\lambda$ before the Planck scale. \\

The above consequences can be significantly altered if the
$\beta$-function of $\lambda$ is modified. Necessarily, physics
beyond the Standard Model needs to be added. There is already a
well-established field of new physics: massive neutrinos. We
therefore analyze the possibility that the physics behind neutrino
mass influences the theoretical Higgs constraints.

The most straightforward mechanism to generate small active neutrino
masses is the type I seesaw
mechanism~\cite{Minkowski:1977sc,*Yanagida:1979as,*Mohapatra:1979ia,*GellMann:1980vs},
in which case neutrino masses are given by \be m_\nu =
\frac{m_D^2}{M_R} = Y_\nu^2 \frac{v^2}{M_R} \, , \ee with $Y_\nu =
m_D/v$ the Yukawa coupling of the neutrinos, $v$ the Standard Model
vacuum expectation value and $M_R$ the mass scale of heavy
right-handed neutrinos. A Yukawa coupling of order one, and a
neutrino mass scale of 0.05 to 0.1 eV, implies for $M_R$ roughly
$10^{14}$ to $10^{15}$ GeV, which defines its canonical scale.
Naively, decreasing $M_R$ indicates smaller Yukawa couplings. If we
run now $\lambda$ from low to high scale, and cross the threshold at
which the heavy neutrinos are integrated out, the Yukawa couplings
of the neutrinos generate the following contribution to the
$\beta$-function of the Higgs
self-coupling~\cite{Grzadkowski:1987tf,*Pirogov:1998tj}:
\begin{eqnarray}
\Delta \beta_\lambda = 4 \lambda {\rm tr} \left(Y^\dagger_\nu Y_\nu
\right) - 8 {\rm tr} \left[  \left( Y^\dagger_\nu Y_\nu \right)^2
\right] .
\end{eqnarray}
We see that the effect of the Dirac Yukawas is very similar to the
one of the top quark, and thus will influence the vacuum stability
bound. One expects that the Higgs mass window will become narrower,
and that the cutoff scale of the theory will decrease
\cite{Casas:1999cd,EliasMiro:2011aa,Gogoladze:2008ak,He:2010sk,Chen:2012fa}.
Sizable effects
 on $\lambda$ imply sizable Yukawa couplings $Y_\nu$, and thus heavy neutrino
masses around $10^{13}\sim10^{15}~{\rm GeV}$. Lighter right-handed
neutrinos reduce the Yukawas and correspondingly their effect on
Higgs mass constraints.

However, in this paper we note that even low scale seesaw models can
have an impact. The simple reason is that the seesaw formula
contains matrices instead of numbers, and thus the argumentation
from above can be avoided: large Yukawa couplings can work very well
with low scale right-handed neutrinos, even going down to the TeV
scale. Note that the scale of the right-handed singlets is a priori
a free and unknown parameter, and might not correspond to the
"canonical" scale of $\gtrsim 10^{14}$ GeV. While this superheavy
scale is expected from the point of view of naturalness, one should
note that also a vanishing right-handed neutrino mass is natural, in
the sense that the symmetry of the Lagrangian is enhanced in that
case. Hence, the scale of $M_R$ is unknown and might be at the
phenomenologically interesting TeV scale. In order to have sizable
mixing of TeV singlets with the SM particles, some tuning is
necessary, which can however be arranged in models. It requires some
peculiar flavor structure in the matrices, and one notes that the
unexpected mixing scheme of leptons seems to hint towards rather
non-trivial flavor structure. This is a framework often applied in
the literature
\cite{Kersten:2007vk,Gavela:2009cd,*Zhang:2009ac,Adhikari:2010yt,*Ohlsson:2010ca,*Ibarra:2011xn,*Haba:2011pe},
and we will discuss the impact on the Higgs mass bounds in this
work.

In this situation described above, the $\beta$-function of $\lambda$
receives sizable corrections even at low energy scales, with
implications on the possible cutoff scale.
 We will give here examples on the consequences of (in particular low scale) type I
seesaw for the evolution of the Higgs self-coupling by using a
general parametrization of the unknown Dirac Yukawas, as well as by
applying two explicit models realizing low scale seesaw with large
Yukawas. We note that sizable effects of neutrino Yukawas on the
Higgs self-coupling can be expected in situations in which
production of right-handed neutrinos
at colliders is possible, because this can only be achieved if the Yukawas are large enough.\\

The remaining parts of this work are organized as follows: In
Sec.~\ref{sec:seesaw} we present the framework of type I seesaw and
its influence on the evolution of the Higgs self-coupling.
Numerical analyses illustrating this will be performed in
Sec.~\ref{sec:numerics}, where we also discuss the impact on the top
Yukawa coupling. Finally, in Sec.~\ref{sec:conclusion}, we summarize
our results and conclude.

\section{Seesaw impact on $\beta_\lambda$} \label{sec:seesaw}

Among various attempts to extend the Standard Model (SM) in order to accommodate
massive neutrinos, the type I seesaw mechanism turns out to be a very
attractive one, in view of its natural and elegant explanation of
light neutrino mass scales. Before studying the vacuum stability in
the seesaw model, we first briefly review the type I seesaw and the
parametrization utilized throughout the remaining parts of this
work.

\subsection{Right-handed neutrinos and seesaw}

In order to generate light neutrino masses, one typically introduces
three right-handed neutrinos besides the Standard Model particle content. The
corresponding Lagrangian reads
\begin{eqnarray}\label{eq:L}
-{\cal L}_\nu =  \overline{\nu_R} Y_\nu \ell_L \tilde H^\dagger +
\frac{1}{2} \overline{\nu_R} M_R \nu^c_R + {\rm h.c.},
\end{eqnarray}
where $\tilde H ={\rm i} \tau_2 H^*$ is the Higgs doublet, and $Y_\nu$ denotes the Yukawa
coupling matrix. At energy scales below the lightest right-handed
neutrino threshold, the light neutrino mass matrix is given by the
well-known seesaw
formula~\cite{Minkowski:1977sc,*Yanagida:1979as,*Mohapatra:1979ia,*GellMann:1980vs}
\begin{eqnarray}\label{eq:m}
m_\nu = v^2 Y^T_\nu M^{-1}_R Y_\nu \; ,
\end{eqnarray}
where $v$ is the vacuum expectation value of $H$. Ignoring the
flavor structures, Eq.~\eqref{eq:m} allows us to naively estimate
the magnitude of $Y_\nu$. For example, for $m_\nu =
{\cal O}(0.1)~{\rm eV}$, one has $Y_\nu$ of order unity if $M_R = {\cal
O}(10^{14})$ GeV. In what follows, we shall consider however the flavor
structure of $Y_\nu$ by using an explicit parametrization scheme.

Adopting the convention defined in Ref.~\cite{Casas:2001sr}, we make
use of the following parametrization of $Y_\nu$,
\begin{eqnarray}
Y_\nu = \frac{1}{v} \sqrt{\bar M_R} R \sqrt{\bar m_\nu} U^\dagger \;
,
\end{eqnarray}
where $\bar M_R$ and $\bar m_\nu$ stand for the diagonal mass
matrices of heavy and light neutrinos, respectively. Here $U$ is a
unitary (PMNS) matrix diagonalizing $m_\nu$ as $\bar m_\nu = U^T m_\nu U
\equiv {\rm diag} (m_1,m_2,m_3)$ with $m_i$ being light neutrino
masses. $R$ denotes a complex orthogonal matrix which we parameterize as
\begin{eqnarray}
R= O e^{{\rm i} A} \; ,
\end{eqnarray}
where both $O$ and $A$ are real matrices. A similar parametrization
has been discussed in a different context in~\cite{Pascoli:2003rq}.
The orthogonality of $R$
implies that $O$ is orthogonal and $A$ is antisymmetric, i.e.,
\begin{eqnarray}
A= \begin{pmatrix} 0 & a & b \cr -a & 0 & c \cr -b & -c & 0
\end{pmatrix} ,
\end{eqnarray}
with real $a,b,c$.
We will show in the following section that such a parametrization
scheme is particularly useful in discussing the evolution of
$\lambda$.

\subsection{Corrections to the RG evolution of $\lambda$}

At energy scales above the right-handed neutrino threshold
additional contributions to the SM $\beta$-functions have to be considered in the evolution of all physical
parameters. For the Higgs self-coupling, $\beta_\lambda$ receives
corrections repeated here for
convenience~\cite{Grzadkowski:1987tf,*Pirogov:1998tj}:
\begin{eqnarray}
\Delta \beta_\lambda = 4 \lambda {\rm tr} \left(Y^\dagger_\nu Y_\nu
\right) - 8 {\rm tr} \left[  \left( Y^\dagger_\nu Y_\nu \right)^2
\right] \; .
\end{eqnarray}
at one-loop level. The first term turns to be small when $\lambda$
approximates to zero, whereas the second term is not suppressed by
$\lambda$ and may affect the RG evolution of $\lambda$ significantly
for large $Y_\nu$. In general, the presence of right-handed
neutrinos drives $\lambda$ towards smaller values at higher energy scales.

For simplicity, we assume that the right-handed neutrino masses are
degenerate, i.e., $M_1=M_2=M_3=M_0$, which allows us to match the
effective theory onto the full theory at a unified scale $\mu=M_0$.
Note that only the combination $Y^\dagger_\nu Y_\nu$ enters the
$\beta$-function of $\lambda$, and one can easily prove that ${\rm
tr}\left(Y^\dagger_\nu Y_\nu\right)$ depends only on the real
parameters in $A$:
\begin{eqnarray}\label{eq:YnuYnu}
{\rm tr}\left(Y^\dagger_\nu Y_\nu\right) & = & {\rm tr} \left[
\frac{M_0}{v^2} U \sqrt{\bar m_\nu} \left(e^{{\rm i}
A}\right)^\dagger e^{{\rm i} A} \sqrt{\bar m_\nu}
U^\dagger \right]= \frac{M_0}{v^2} {\rm tr} \left[  \sqrt{\bar m_\nu}   e^{2{\rm i}
A} \sqrt{\bar m_\nu} \right]  \\
&=&  \frac{M_0}{v^2} \left[ \sum_i m_i + \frac{4(\cosh r-1)}{r^2}
\left( m_1 (a^2+b^2) + m_2 (a^2+c^2) + m_3 (b^2+c^2)
\right)\right] \nonumber,
\end{eqnarray}
where $r=2\sqrt{a^2+b^2+c^2}$. One observes that neither the
rotation matrix $O$ nor the mixing matrix $U$ affects the evolution
of $\lambda$ directly. Furthermore, the hyperbolic cosine factor
involving $r$ plays a crucial role in seesaw models since it could
enhance the magnitude of $Y_\nu$ without spoiling the values of the light neutrino
masses. In other words, the $R$ matrix elements can be arbitrarily
large in virtue of the fact that $R$ disappears from the seesaw
formula playing no role in determining the light neutrino
masses\footnote{Obviously, the elements of $Y_\nu$ need to be
perturbative.}.
However, $R$ does influence the Yukawa couplings in $Y_\nu$ which could
lead to sizable impact on the running of $\lambda$.

In case that the light neutrino masses are quasi-degenerate, namely
$m_1 \simeq m_2 \simeq m_3 = m_0$, Eq.~\eqref{eq:YnuYnu} can further be
simplified to
\begin{eqnarray}\label{eq:YnuSim}
{\rm tr}\left(Y^\dagger_\nu Y_\nu\right) \simeq \frac{M_0 m_0}{v^2}
\left(1+ 2 \cosh r\right)  .
\end{eqnarray}
As a numerical example, we take $M_0 = 1~{\rm TeV}$, $m_0
=0.1~{\rm eV}$, $r=25$, and obtain ${\rm tr}\left(Y^\dagger_\nu
Y_\nu\right) = {\cal O}(0.1) $. The naive estimate from $m_\nu =
Y_\nu^2 v^2/M_R $ would give $Y_\nu^2 \sim 10^{-11}$ and thus a
negligible effect. Therefore, even for very light right-handed
neutrinos, sizable Yukawa couplings are still acceptable when the
matrix structure of $Y_\nu$ is taken properly into account.
Arranging sizable Dirac Yukawas in case of a low seesaw scale is
often done in order to construct scenarios in which the seesaw
messengers are to be produced at colliders. The reason is that the
mixing with the SM particles is given by $S \sim Y_\nu v/M_R$, which
needs to be not too small in order to allow sizable cross
sections~\cite{Datta:1993nm,Han:2006ip,Aguila:2007em}. As a further
illustration of this, consider the case in which the mixing of one
right-handed neutrino with SM particles, say, electrons, saturates
the present bound~\cite{Bergmann:1998rg,*Bekman:2002zk}: $|S_{ei}|^2
\le 0.0054$. This implies that $Y_\nu \simeq 0.4$ $(M_R/$TeV). We
stress that if right-handed neutrinos can be produced at colliders,
the effects on the Higgs self-coupling and electroweak vacuum
stability that we discuss in this paper can be expected.

We note that the above analytical approximations rely on the
assumption of a degenerate (or nearly degenerate) right-handed
neutrino mass spectrum. Relaxing this assumption will not allow us to
eliminate the rotation matrix
$O$ in Eq.~\eqref{eq:YnuYnu}, and thus increase the number of free
parameters (three real rotation angles are necessary to parameterize
$O$). It will furthermore complicate the analysis significantly, as
several thresholds would have to be taken into account. In analogy to
Refs.~\cite{Casas:1999cd,EliasMiro:2011aa,Chen:2012fa}, we will stick
to degenerate heavy neutrinos, but in contrast to those works (which
used $R = \mathbbm 1 $) take the
flavor structure of the Yukawa matrix into account.

Finally we stress that the complexity of $e^{{\rm i}A}$ is crucial
in the leptogenesis mechanism~\cite{Fukugita:1986hr}. The presence
of light right-handed neutrinos may also have impact on various low
scale phenomena, e.g., lepton flavor violating processes,
non-unitarity effects in neutrino oscillations, neutrinoless double
beta decay as well as signatures at colliders.

\section{Numerical illustrations} \label{sec:numerics}

We proceed to illustrate the previous discussion on how the
right-handed neutrinos affect the evolution of $\lambda$. In our
numerical analysis, we make use of the input physical parameters
(e.g., fermion masses, gauge couplings, flavor mixing parameters)
from recent re-evaluations of running SM parameters normalized at
the electroweak scale, i.e.,
$\mu=M_Z$~\cite{Xing:2007fb,Xing:2011aa}. The full set of two-loop
RGE~\cite{Machacek:1983tz,*Machacek:1983fi,*Machacek:1984zw,*Arason:1991ic,*Ford:1992mv,*Luo:2002ey}
for $\lambda$ is solved together with the one-loop matching
condition for $\lambda(\mu)$ and
$m_t$~\cite{Sirlin:1985ux,*Hambye:1996wb}. For simplicity, we also
assume $a=b=c=a_0$, implying that $r=2\sqrt{3} a_0$. In addition,
only the normal mass ordering ($m_1<m_2<m_3$) of light neutrinos is
considered; for a nearly degenerate neutrino mass spectrum there is
no quantitative difference between normal and inverted mass
orderings. For simplicity, and in order to focus on the impact of
the neutrino sector, we will use the top quark pole mass $m^{\rm
pole}_t = 172.9~{\rm GeV}$ advocated by the Particle Data
Group~\cite{Nakamura:2010zzi}, and will not take into account the
effect of varying the top quark mass or the strong coupling.

\subsection{Corrections to the Higgs self-coupling}

\begin{figure}[t]
\begin{center}
\includegraphics[width=.49\textwidth]{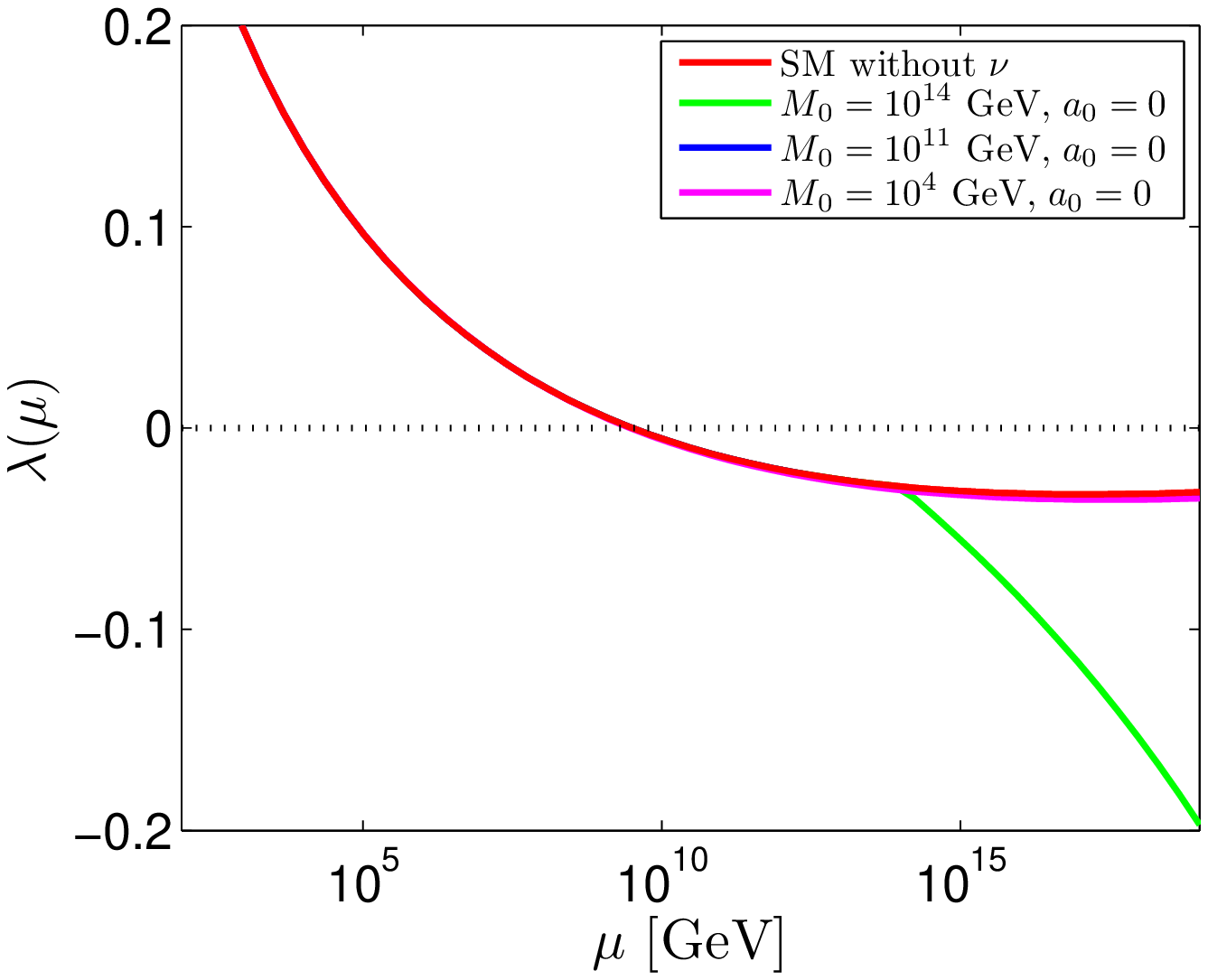}
\includegraphics[width=.49\textwidth]{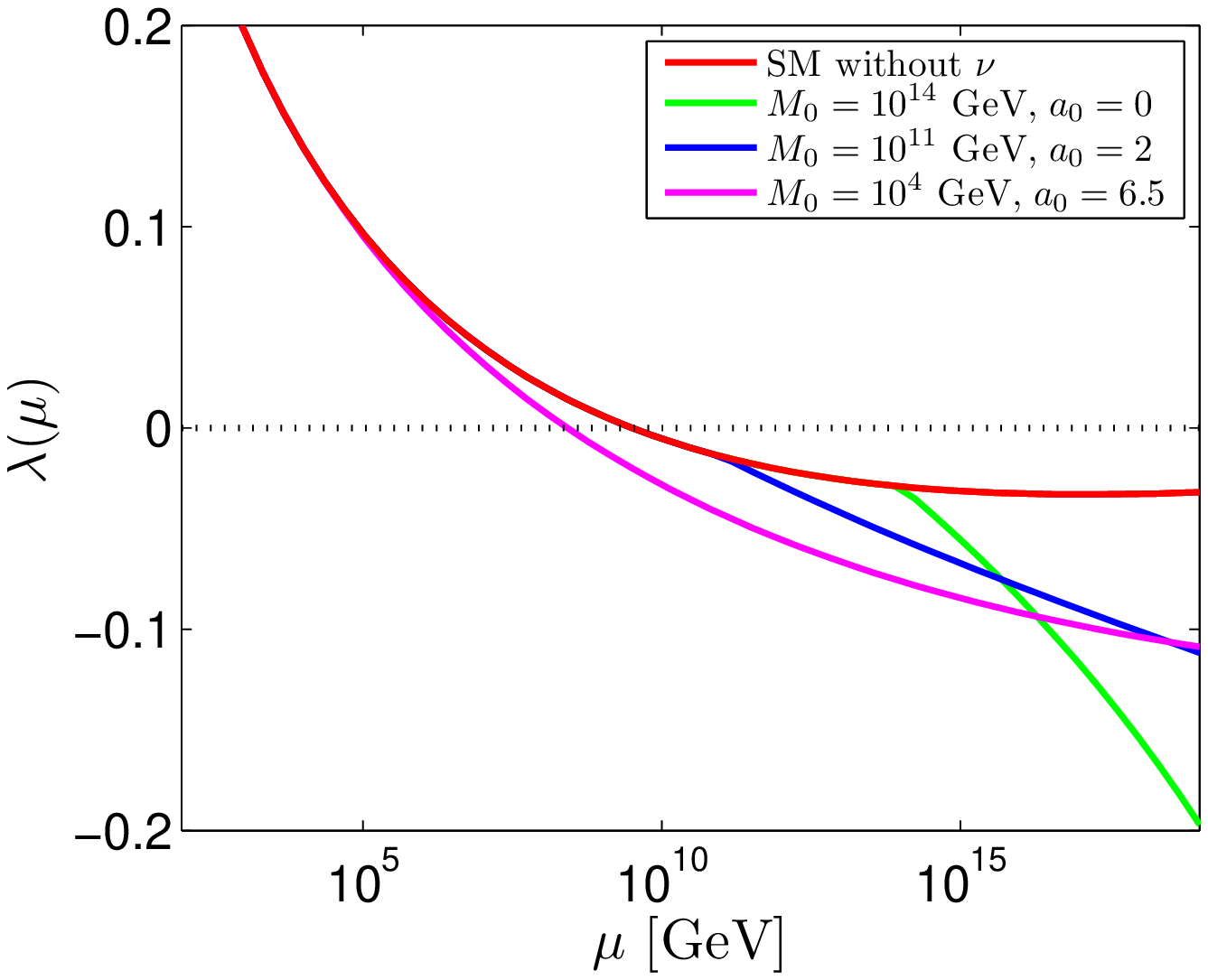}
\includegraphics[width=.49\textwidth]{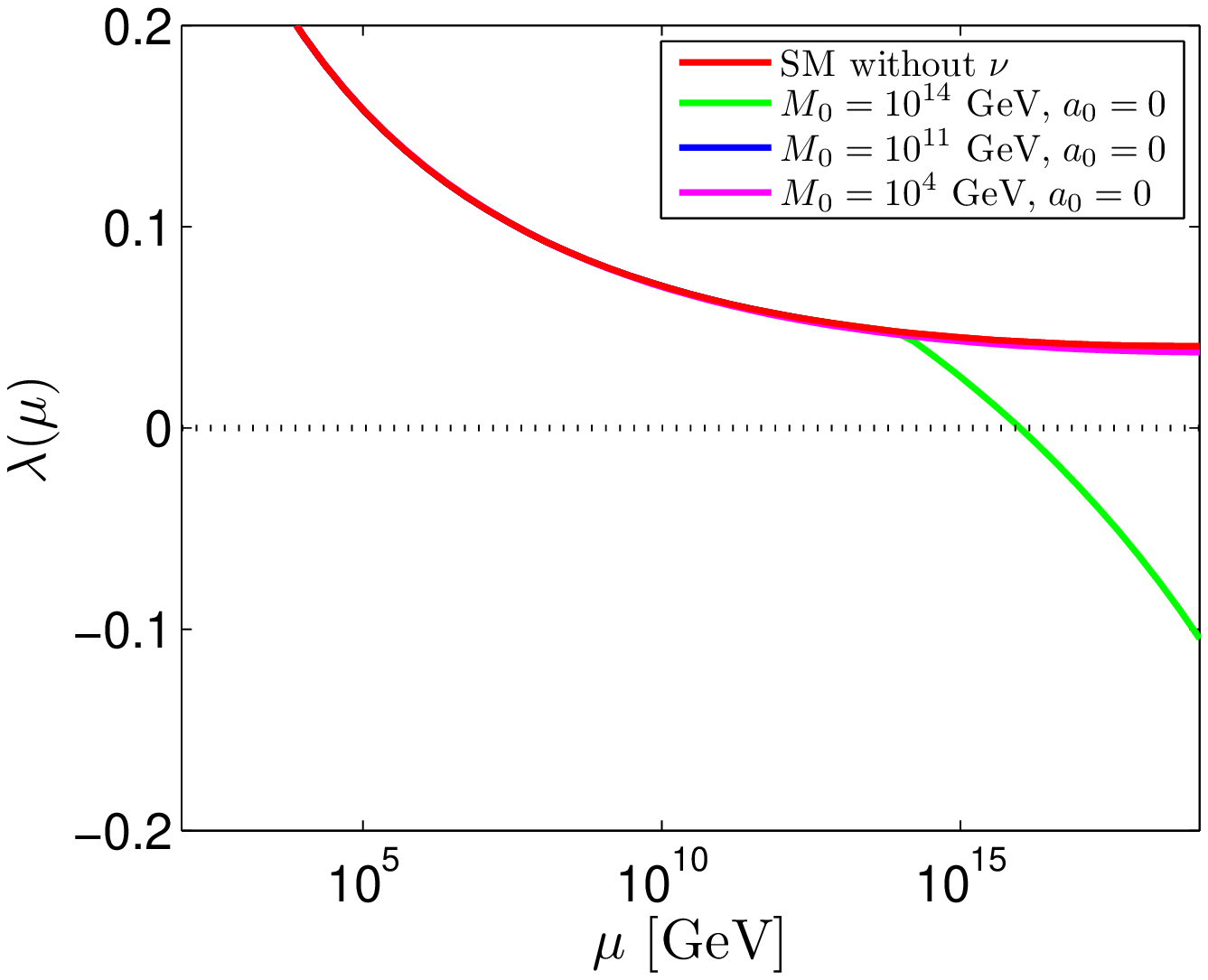}
\includegraphics[width=.49\textwidth]{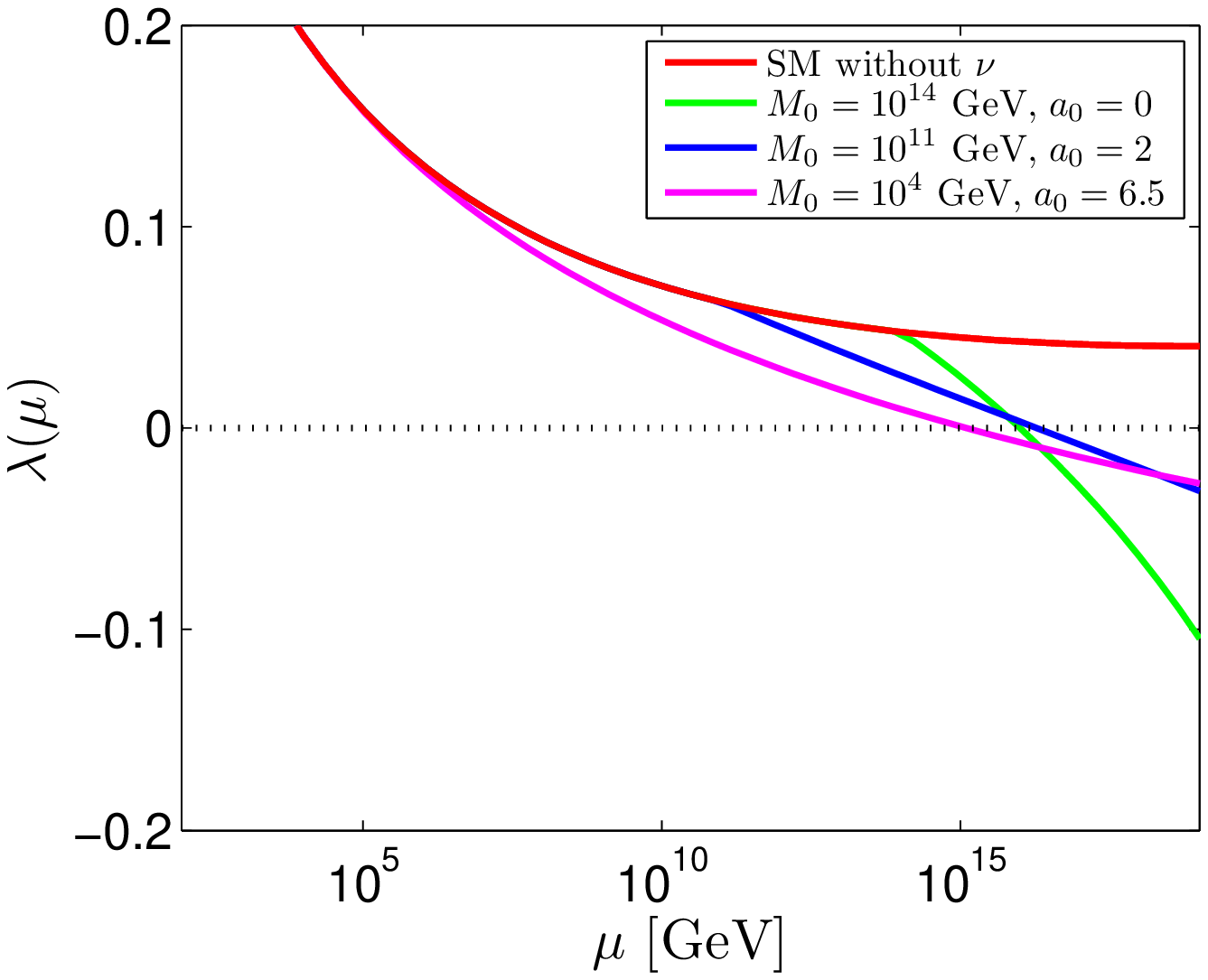}
\caption{ \label{fig:fig_lambda} RG evolution of the Higgs
self-coupling for $m_H=125~{\rm GeV}$ (upper panel) and $m_H=135~{\rm
GeV}$ (lower panel) with $m_1=0.1~{\rm eV}$. Red curves correspond
to the SM without right-handed neutrinos, while for green
and blue curves we include right-handed neutrino contributions and
take the right-handed neutrino thresholds $M_0=10^{14}~{\rm GeV}$
(green), $M_0=10^{11}~{\rm GeV}$ (blue) and $M_0=10~{\rm TeV}$
(magenta). In the left plots we switch
off the $A$ matrix by setting $a_0 =0$, whereas for the right plots,
we take $a_0=2$ and $a_0=6.5$ for the green and blue curves,
respectively.}
\end{center}
\end{figure}

In Fig.~\ref{fig:fig_lambda} we show the impact of right-handed
neutrinos on the evolution of $\lambda(\mu)$. For illustration
purpose, we take $m_H=125~{\rm GeV}$ and $m_H=135~{\rm GeV}$
together with the light neutrino mass $m_1=0.1~{\rm eV}$. The value
125 GeV corresponds to a situation in which even in the SM there is
a non-trivial cutoff scale, whereas 135 GeV is for our parameter
choice a value which in the
absence of SM extensions causes no conceptual issues up to the
Planck scale. The red curves in Fig.~\ref{fig:fig_lambda} show the
running behavior in the pure SM case, from which one can see that
for $m_H=125~{\rm GeV}$ the self-coupling $\lambda$ crosses zero at
energy scales around $10^{10}~{\rm GeV}$, indicating an upper bound
on the potential new physics scale. In case of $m_H=135~{\rm GeV}$,
the SM is valid in principle up to the Planck scale. The green and
blue curves show the running of $\lambda$ with right-handed neutrino
mass scale $M_0=10^{11}~{\rm GeV}$ and $M_0=10~{\rm TeV}$,
respectively. For the left plots, we switch off the $A$ matrix by
setting $a_0=0$. As can be seen from the plot, the evolution of
$\lambda$ is barely affected by right-handed neutrinos (the curves
almost overlap with each other), which is consistent with our
analytical analysis since $Y_\nu$ is small in order to compensate a
light right-handed neutrino scale. However, when we switch on the
real parameters in the $A$ matrix, the evolution of $\lambda$ is
modified significantly after crossing the seesaw threshold. Even for
relatively light right-handed neutrinos, one can see clearly the
shift of the cutoff scale down to low energy ranges. Noteworthy is
for instance that the case $m_H=135~{\rm GeV}$ (which is safe in the
SM) can have an unstable vacuum. This can happen even in the
canonical case without any non-trivial $R$, as can be seen by the
green line corresponding to $M_0= 10^{14}$~GeV~\cite{Casas:1999cd}.

\subsection{Higgs mass window and cutoff scale}

\begin{figure}[t]
\begin{center}
\includegraphics[width=.6\textwidth]{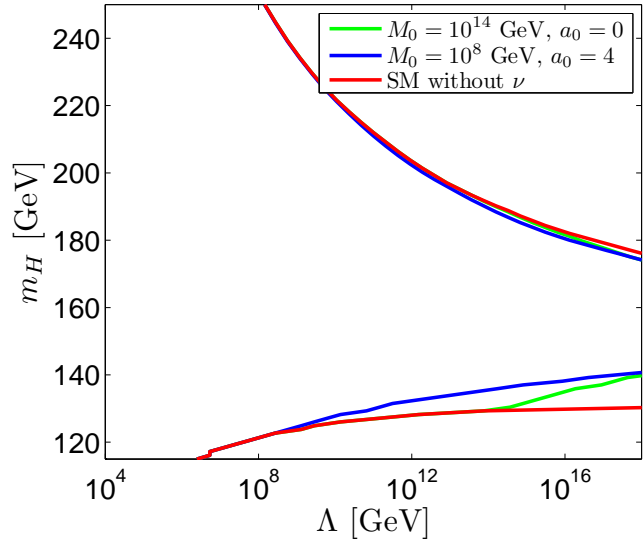}
\caption{ \label{fig:fig_window} Higgs mass window from
stability and triviality bounds. The red curve corresponds to the case
without right-handed neutrinos, while for green and blue curves we
take ($M_0=10^{14}~{\rm GeV},~a_0=0$) and ($M_0=10^{8}~{\rm
GeV},~a_0=4$), respectively.}
\end{center}
\end{figure}

The effect of right-handed neutrinos on the stability bounds is
shown in Fig.~\ref{fig:fig_window} for particular choices of
parameters. The threshold effects take place at $\mu=M_0$ and can be
identified easily on the plot. In the SM framework, there is
essentially no stability constraint for Higgs masses larger than
about $130~{\rm GeV}$, whereas in the presence of right-handed
neutrinos the cutoff scale $\Lambda$ is decreased significantly, in
particular for low scale seesaw models with large $A$ matrix
elements. For completeness, we also show in
Fig.~\ref{fig:fig_window} the triviality bounds on the Higgs mass
improved by including right-handed neutrinos, i.e., $\lambda(\mu)
<4\pi$. As expected, there is very little effect from seesaw. One
can read from the plot that the Higgs mass window becomes narrower
when the right-handed neutrinos are added, which is mainly due to
the strengthening of the stability bounds.

\begin{figure}[t]
\begin{center}\vspace{-0.25cm}
\includegraphics[width=.7\textwidth]{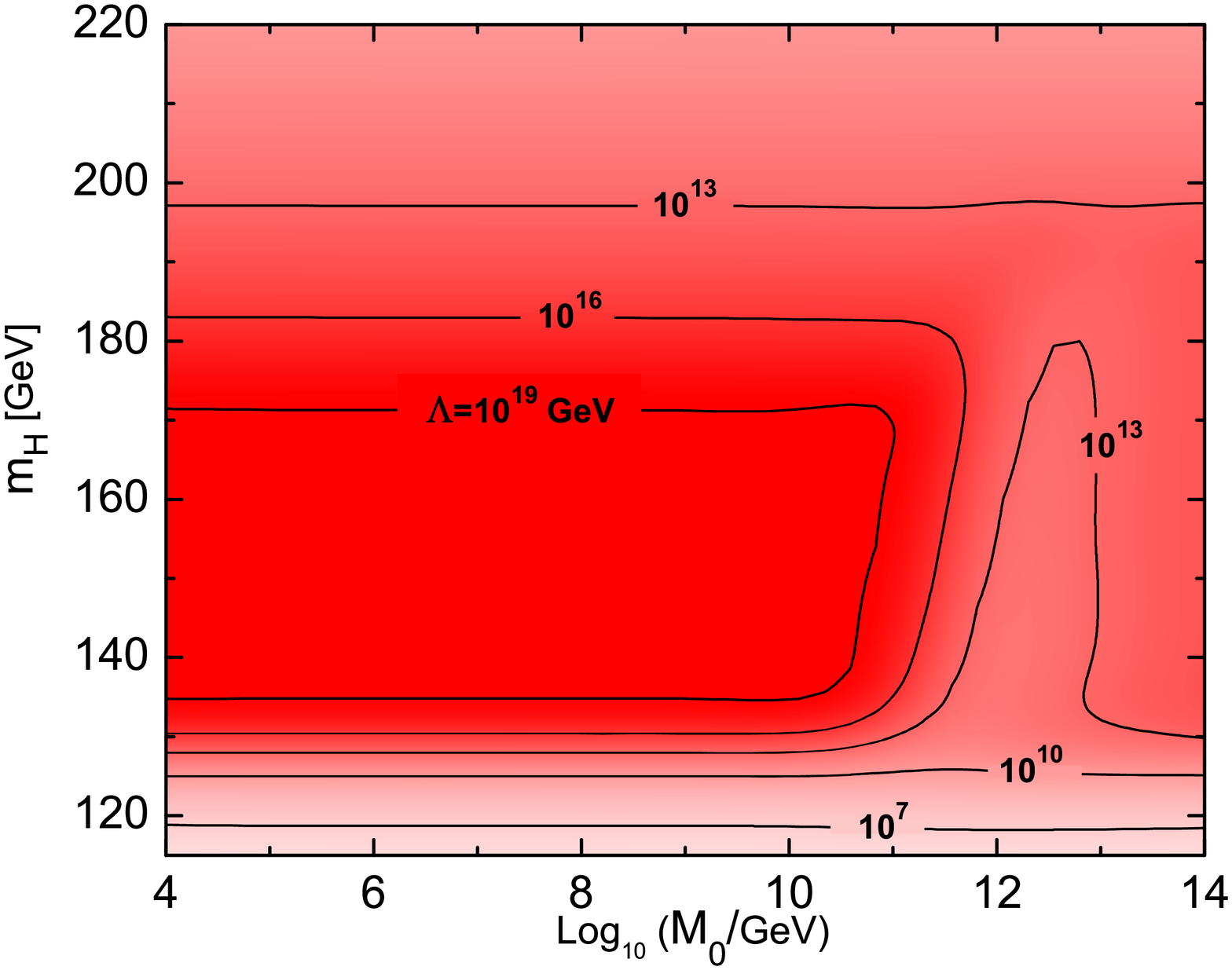}
\caption{ \label{fig:fig_M0} Illustration for the parameter space of
$m_H$ and $M_0$, for different choices of the cutoff scale
$\Lambda$. Darker areas correspond to higher cutoff scales allowed
by the stability and triviality boundary conditions. The other input
parameters are taken as $m_1=0.1~{\rm eV}$ and
$a_0=2$.}\vspace{-0.3cm}
\end{center}
\end{figure}
\begin{figure}[t]
\begin{center}\vspace{-0.7cm}
\includegraphics[width=.7\textwidth]{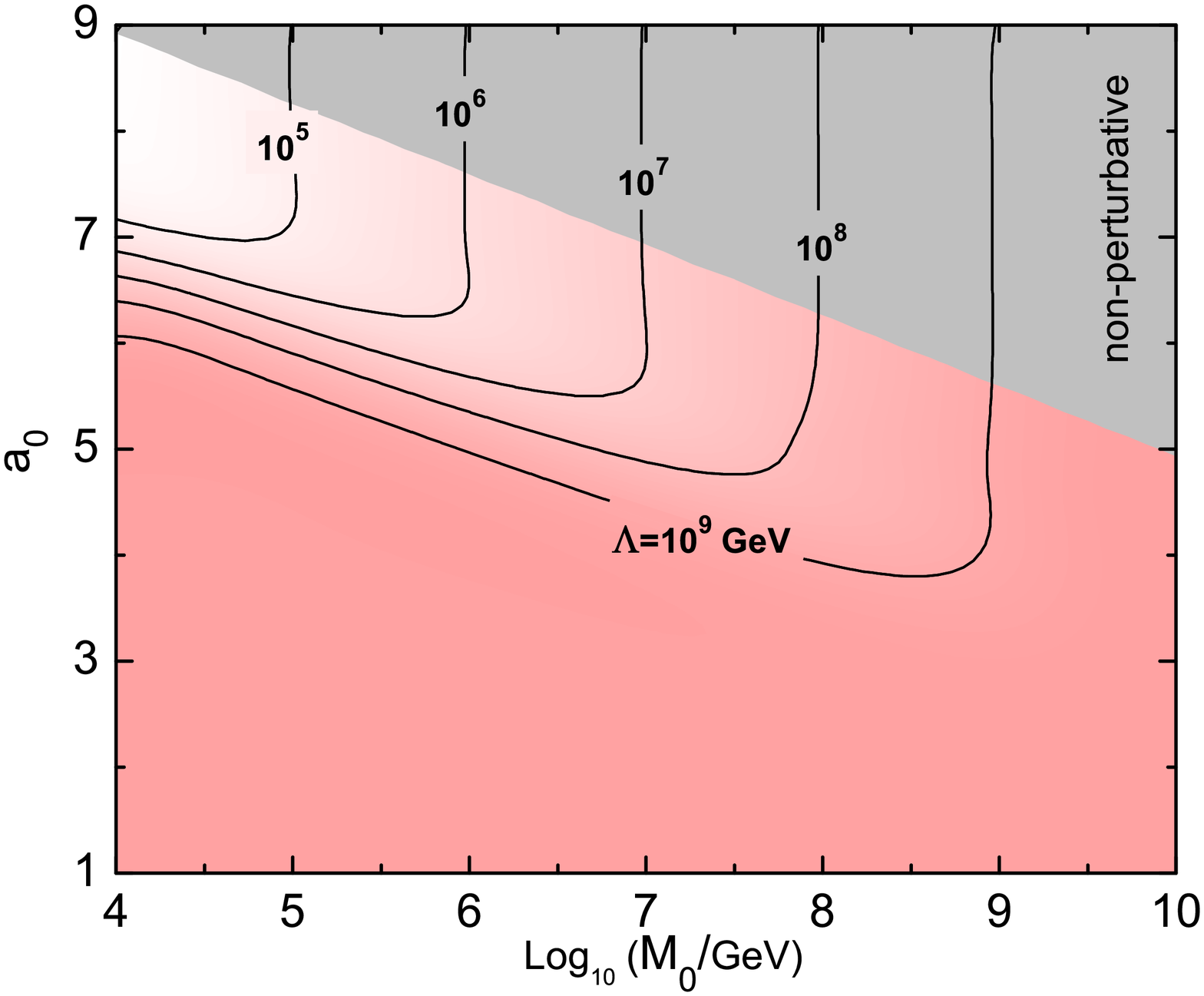}
\includegraphics[width=.7\textwidth]{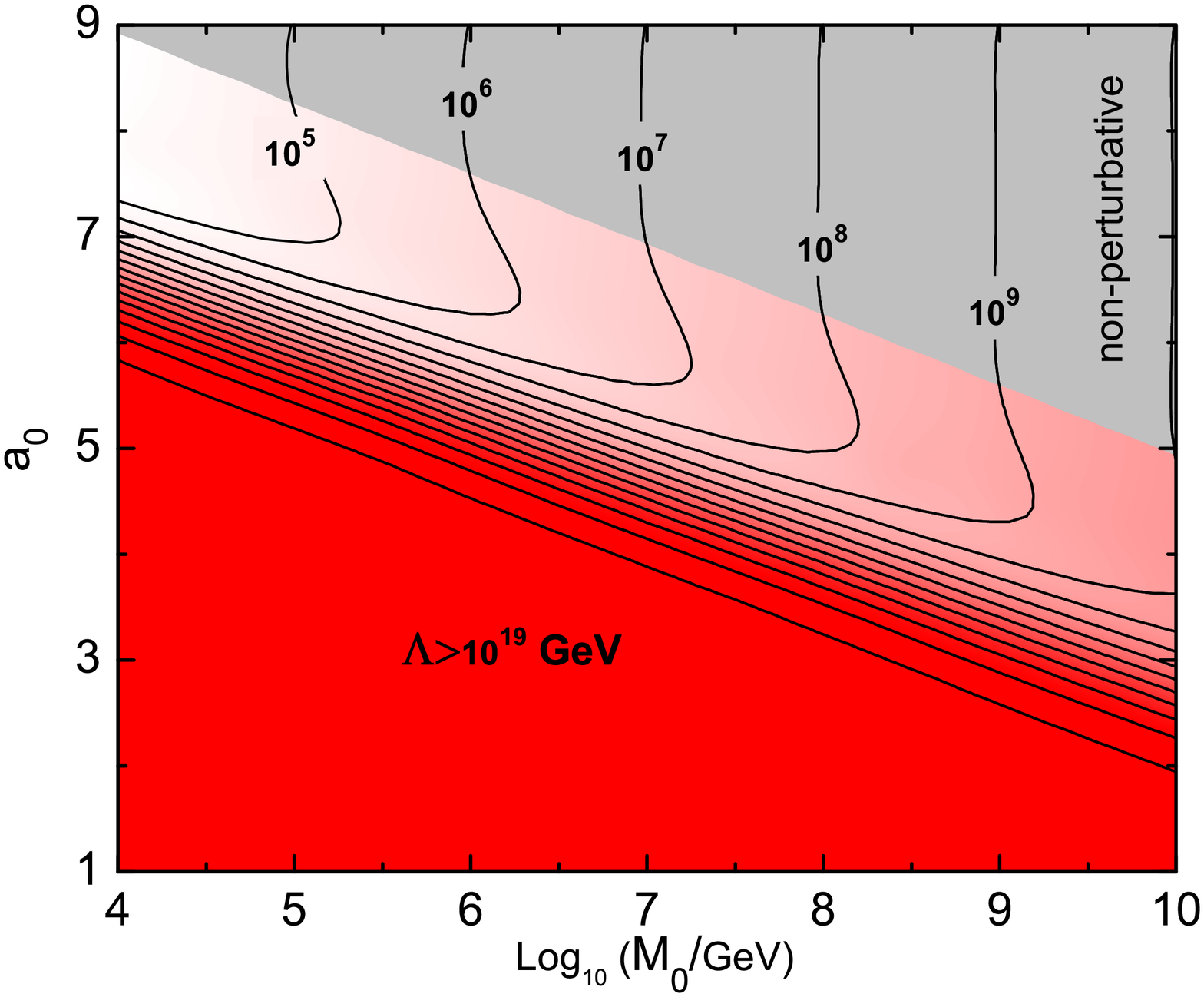}
\caption{\label{fig:fig_a0} Parameter ranges of $M_0$ and $a_0$ for $m_1 = 0.1$ eV,
$m_H =125~{\rm GeV}$ (upper plot) and $m_H=135~{\rm GeV}$ (lower
plot). The areas shaded in gray correspond to non-perturbative
couplings, i.e., the absolute value of one entry in $Y_\nu$ is
larger than $4\pi$.}
\end{center}
\end{figure}

As mentioned above, in determining the cutoff scale of possible new
physics, the neutrino Yukawas can be crucial and may affect
$\beta_\lambda$ in a similar way as the top quark mass. We
illustrate in Fig.~\ref{fig:fig_M0} the allowed parameter ranges in
the $m_H$--$M_0$ plane by choosing the light neutrino mass
$m_1=0.1~{\rm eV}$ and $a_0=2$. One can read from the plot that, for
a chosen Higgs mass smaller than $125~{\rm GeV}$, the cutoff scale
is essentially not affected by right-handed neutrinos (cf.~the
horizontal lines at the bottom of Fig.~\ref{fig:fig_M0}), since the
corresponding Yukawa couplings are not significant enough when
right-handed neutrino thresholds are lighter than $10^{10}~{\rm
GeV}$. The situation becomes more complicated for a Higgs mass in
the intermediate range, i.e., $ 130~ {\rm GeV}\lesssim m_H<180
\lesssim {\rm GeV}$, owing to the fact that there is essentially no
stability constraint in the pure SM framework (i.e., $\Lambda$ can
be as large as the Planck scale $10^{19}~{\rm GeV}$). For example,
if the Higgs mass is located at $m_H=140~{\rm GeV}$, the cutoff
scale $\Lambda$ decreases with increasing $M_0$, because $Y_\nu$
becomes larger in this case. When $\Lambda \gtrsim M_0$, $\Lambda$
stops decreasing and achieves a minimum, which can be seen from the
plot for right-handed neutrino masses around a critical scale
$M^{\rm cric}_0\sim 10^{12}~{\rm GeV}$. This can be easily
understood because $\Lambda$ cannot be smaller than $M_0$, otherwise
the right-handed neutrino threshold will not be
crossed.\footnote{Note that a negative $\lambda$ arises for $\Lambda
< M_0$, which, however, does not necessarily mean that the model is
invalid, since the unlikely possibility of a metastable electroweak
vacuum exists.  For a general discussion on this topic, see, e.g.
Ref.~\cite{Sher:1988mj}.} Then, for right-handed neutrino masses in
the range $M_0
> M^{\rm cric}_0$, the RG evolution between $M_0$ and $\Lambda$ is
strongly enhanced by larger $Y_\nu$, which drives $\lambda$ to zero
very efficiently, even for a small energy interval. Therefore, the
cutoff $\Lambda$ will remain close to $M_0$, which is reflected in
Fig.~\ref{fig:fig_M0} by the vertical parts of the contour $\Lambda
= 10^{13}~{\rm GeV}$. Finally, when $m_H>180~{\rm GeV}$, $\Lambda$
is mainly constrained by the triviality bounds (cf.~the horizontal
lines at the top of Fig.~\ref{fig:fig_M0}). We should note that the
features discussed here do not rely on the specific choice of model
parameters. In general, the value of  $M^{\rm cric}_0$ decreases
with decreasing $a_0$.

We also show in Fig.~\ref{fig:fig_a0} the parameter space of $M_0$
and $a_0$ constrained by the stability bound for $m_H =125~{\rm
GeV}$ and $135$ GeV. From the plot one observes that the role of
right-handed neutrinos becomes visible in case of a relatively large
$a_0$ (e.g., $a_0>4$). In general, for a chosen $M_0$, the larger
$a_0$ the smaller the cutoff scale one can expect. Interestingly,
when both $M_0$ and $a_0$ are sizable, $\Lambda$ remains close to
the right-handed neutrino masses ($\Lambda \simeq M_0$), due to the
same reason mentioned in the above paragraph. This feature is shown
by the vertical curves.

\begin{figure}[t]
\begin{center}
\includegraphics[width=.49\textwidth]{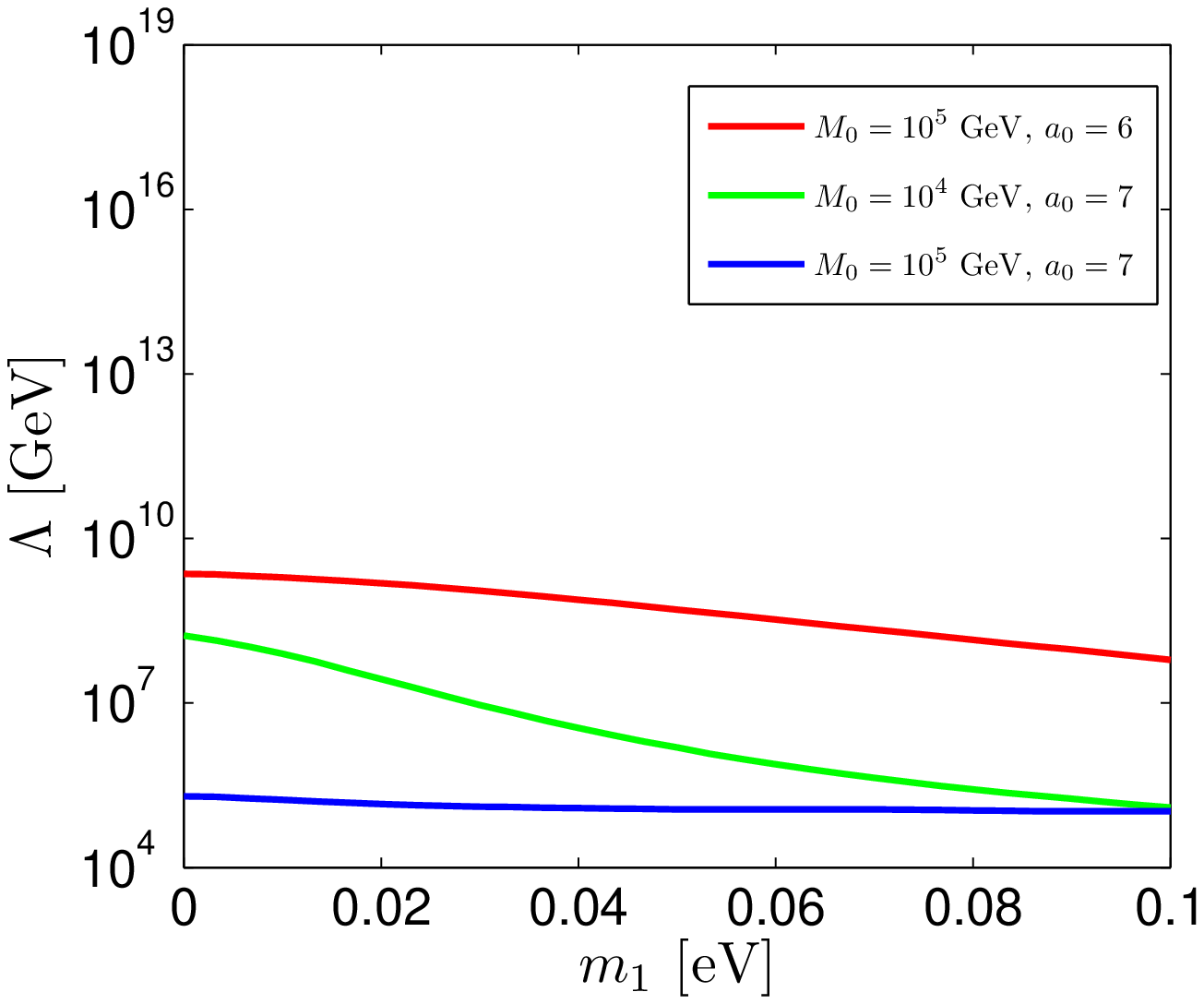}
\includegraphics[width=.49\textwidth]{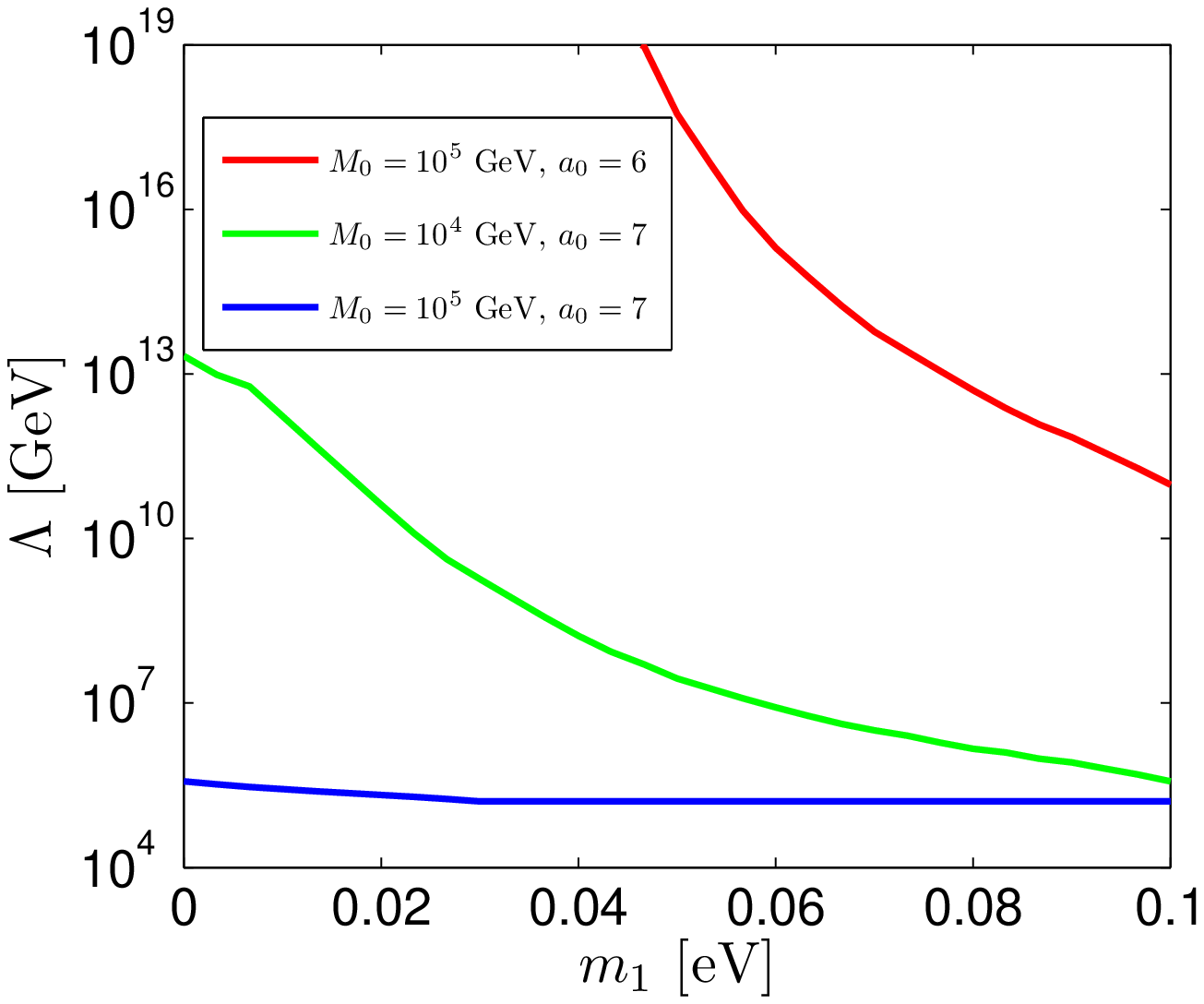}
\caption{ \label{fig:fig_m1} Cutoff scale $\Lambda$ with respect
to the lightest neutrino mass $m_1$ for $m_H=125~{\rm GeV}$ (left
plot) and $m_H=135~{\rm GeV}$ (right plot). Red, green, and blue
curves correspond to the parameter choices ($M_0=10^{5}~{\rm
GeV},~a_0=6$), ($M_0=10^{4}~{\rm GeV},~a_0=7$) and
($M_0=10^{5}~{\rm GeV},~a_0=7$), respectively.}
\end{center}
\end{figure}

The RG evolution of $\lambda$ depends also on the light neutrino
masses as can be seen in Eq.~\eqref{eq:YnuYnu}. Thus, we illustrate
in Fig.~\ref{fig:fig_m1} the dependence of $\Lambda$ on the lightest
neutrino mass $m_1$. One can observe from the plot that, for a
chosen right-handed neutrino scale, the larger the light neutrino
mass, the lower the value at which $\Lambda$ could be located. This
is in good agreement with our analytical results
[cf.~Eqs.~\eqref{eq:YnuYnu} and \eqref{eq:YnuSim}]. Moreover,
$\Lambda$ is very sensitive to $a_0$, in particular when $a_0$ is
large. Note however that $a_0$ cannot be arbitrarily large since
$Y_\nu$ suffers from the perturbativity constraint. Furthermore,
when right-handed neutrinos are relatively light (e.g.~$\lesssim
10~{\rm TeV}$), the unitarity of the PMNS matrix $U$ sets
constraints on the mixing between light and heavy neutrinos,
suggesting roughly $(Y_\nu v /M_0 )^2 \ls 10^{-3}$~\cite{Bergmann:1998rg,*Bekman:2002zk}.

\subsection{RG evolution of $\lambda$ in realistic low scale seesaw models}
While the discussion made so far used a general parametrization of
the unknown neutrino Yukawas, specific models or frameworks
constructed in order to predict certain features of lepton mixing
will be able to provide more definite information on the flavor
structure. We will shortly discuss two examples. Both of them were
constructed in order to have sizable cross sections for the
production of heavy neutrinos at the LHC. \\

\begin{figure}[t]
\begin{center}
\includegraphics[width=.49\textwidth]{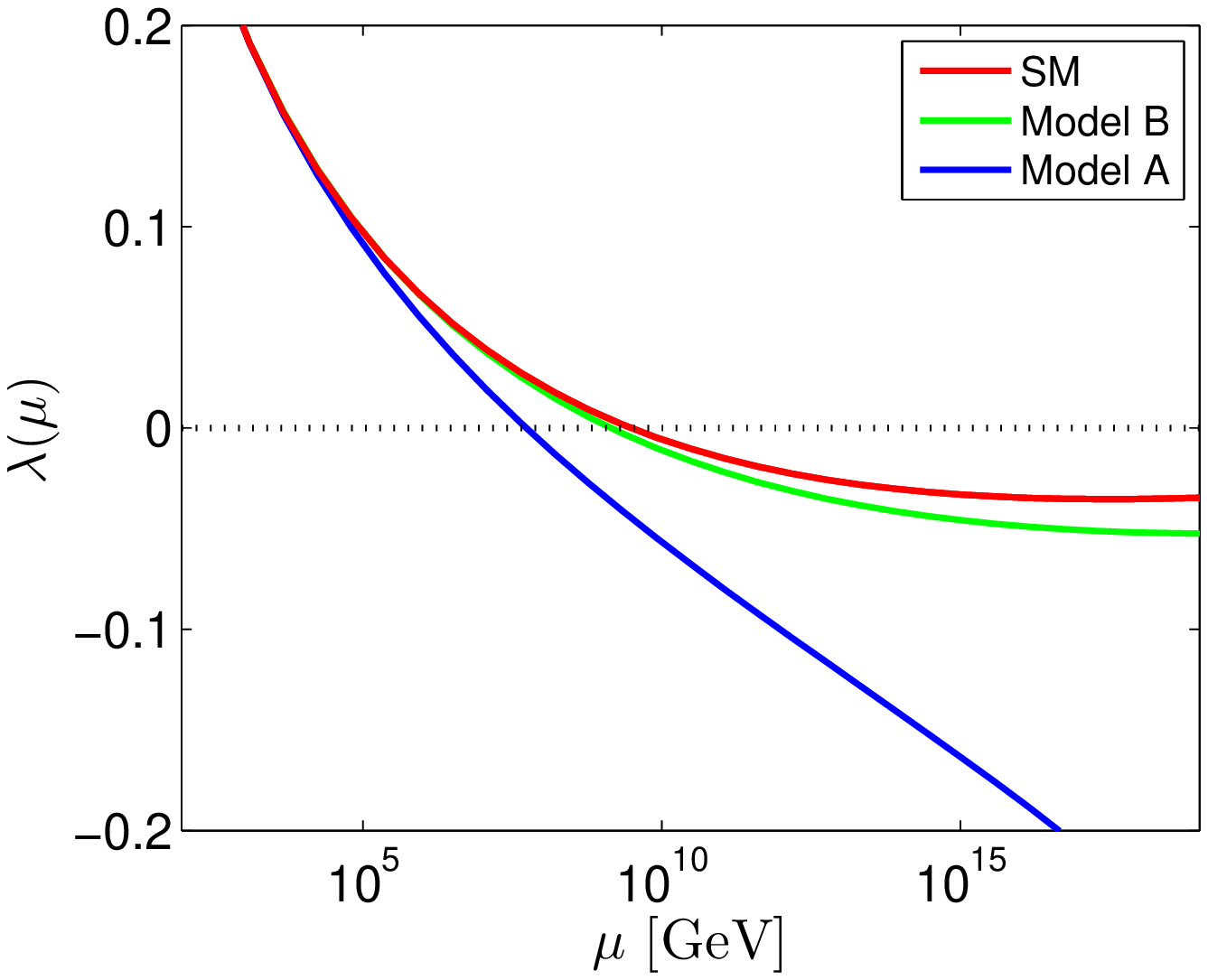}
\includegraphics[width=.49\textwidth]{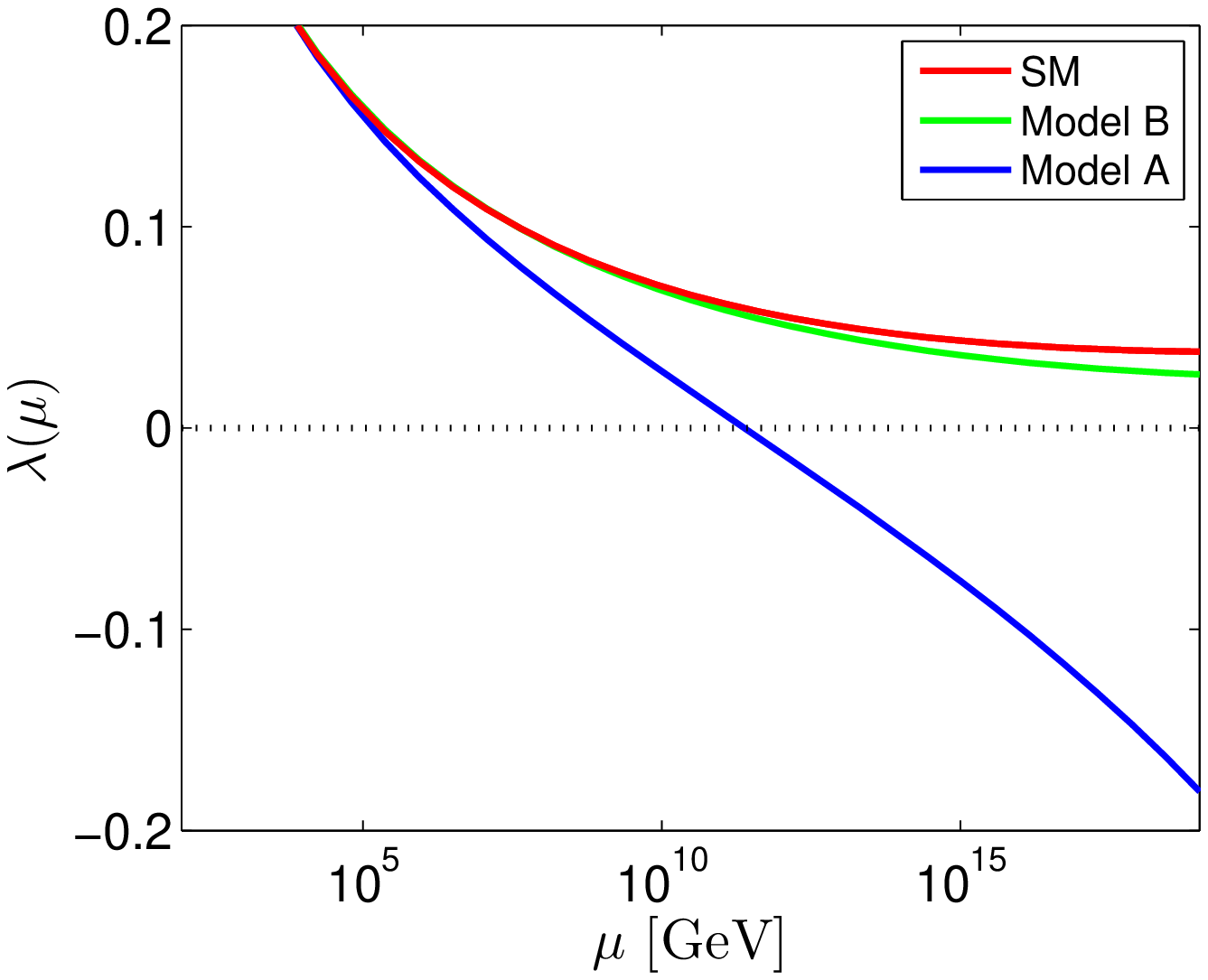}
\caption{ \label{fig:fig_models} RG evolution of the Higgs
self-coupling for $m_H=125~{\rm GeV}$ (left plot) and $m_H=135~{\rm
GeV}$ (right plot). Red curves correspond to the simplest SM without
right-handed neutrinos, while for the green (model A) and blue
(model B) curves, we take the right-handed neutrino threshold
$M_0=350~{\rm GeV}$. The Yukawa coupling parameters are chosen to be
$h_1=h_2=h_3=0.15$ and $y_1=y_2=y_3=0.15$, respectively.}
\end{center}
\end{figure}

{\bf Model A}: We first consider an $A_4$ type I seesaw model
described in Ref.~\cite{Kersten:2007vk}, in which right-handed
neutrinos are assigned to the three-dimensional representation of
$A_4$ while lepton doublets transform under ${\bf 1}''$. At leading
order, $Y_\nu$ is given by
\begin{eqnarray}
Y_\nu = \begin{pmatrix} h_1 & h_2 & h_3 \cr \omega h_1  & \omega h_2
& \omega h_3 \cr \omega^2 h_1  &\omega^2  h_2  & \omega^2 h_3
\end{pmatrix}  ,
\end{eqnarray}
with $\omega = e^{{\rm i}\frac{2}{3}\pi}$, and the right-handed
neutrino mass matrix is proportional to a unit matrix, i.e., $M_R
=M_0 \mathbbm{1}$. Using the seesaw formula \eqref{eq:m}, one can easily
see that light neutrinos are massless in the leading order
approximation. Small perturbations or explicit symmetry breaking terms are
therefore needed to include neutrino masses. We can ignore them
for the purpose of our study.
The neutrino contribution to $\beta_\lambda$ is estimated by
\begin{eqnarray}
{\rm tr}\left(Y^\dagger_\nu Y_\nu\right) \simeq 3 \left(h^2_1+ h^2_2
+ h^2_3 \right)  .
\end{eqnarray}
As an example, we take $h_1=h_2=h_3=0.15$ together with
$M_0=350~{\rm GeV}$, for which a discovery search can be performed
at the LHC~\cite{Aguila:2007em}. We illustrate in
Fig.~\ref{fig:fig_models} the effect of right-handed neutrinos on
the RG evolution of $\lambda$. Similar to Fig.~\ref{fig:fig_lambda},
the right-handed neutrino corrections are
noteworthy compared to the SM evolution.\\

{\bf Model B}: Next we consider the minimal TeV seesaw
model~\cite{Gavela:2009cd,*Zhang:2009ac}, in which only two
right-handed neutrinos are introduced, having opposite CP parity. In
this model, right-handed neutrinos can be paired together to form a
pseudo-Dirac fermion, and light neutrinos are massless at leading
order due to the conservation of lepton number. Effectively, at
leading order, the Yukawa coupling matrix is given by
\begin{eqnarray}
Y_\nu = \begin{pmatrix} y_1 & y_2 & y_3 \cr 0 & 0 & 0
\end{pmatrix} ,
\end{eqnarray}
while the right-handed mass matrix takes a $2\times 2$ form,
\begin{eqnarray}
M_R = \begin{pmatrix} 0 & M_0 \cr M_0 & 0
\end{pmatrix}  .
\end{eqnarray}
It is not difficult to calculate that\footnote{Note that
diagonalization of $M_R$ is not necessary, because the unitary
matrix $V$ that diagonalizes it will appear together with $Y_\nu$ as
$V Y_\nu $ and thus drops out of $Y^\dagger_\nu Y_\nu$.}
\begin{eqnarray}
{\rm tr}\left(Y^\dagger_\nu Y_\nu\right) \simeq y^2_1 + y^2_2 +
y^2_3\; .
\end{eqnarray}
The running of $\lambda$ for the parameter choice $y_1=y_2=y_3=0.15$
is shown in Fig.~\ref{fig:fig_models}. In comparison to model A, a
factor $3$ is missing in ${\rm tr}\left(Y^\dagger_\nu Y_\nu\right)$,
and therefore the modification to the running of $\lambda$ is
milder.

Note that, in order to accommodate non-vanishing neutrino masses in
those models, small lepton number violating perturbations have to be
introduced
\cite{Kersten:2007vk,Gavela:2009cd,*Zhang:2009ac,Adhikari:2010yt,*Ohlsson:2010ca,*Ibarra:2011xn,*Haba:2011pe},
whose stability has to be guaranteed by symmetries. Explicitly, in
order to generate sub-eV light neutrino masses, one needs
perturbation terms $\epsilon$ in the Yukawa matrix of the order
$\epsilon \sim \frac{m_\nu M_R}{y_i v^2}$. Taking Model B as an
example, we can estimate that $\epsilon \sim 10^{-11}$. The
mechanism behind this stability and the precise values of the
parameters do not affect the main results of this work.

\subsection{Top quark Yukawa coupling}

\begin{figure}[t]
\begin{center}
\includegraphics[width=.6\textwidth]{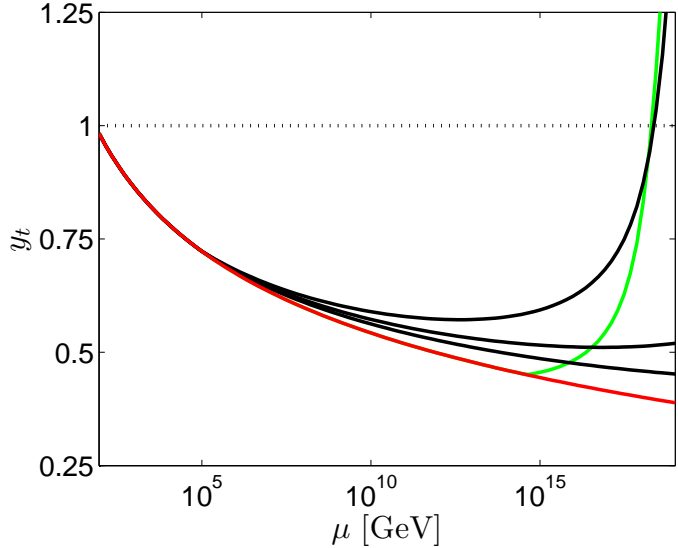}
\caption{ \label{fig:fig_top} RG evolution of top quark Yukawa
coupling $y_t$. We take the input parameters $m_H=125~{\rm GeV}$ and
$m_1=0.1~{\rm eV}$. The red curve corresponds to the pure SM case
without right-handed neutrinos, while the green curve is for
$M_0=4\times 10^{14}~{\rm GeV}$ and $a_0=0$. The black curves
correspond to $M_0=10^5~{\rm GeV}$ with $a_0=6.0,~6.1,~6.2$ from
bottom to top.}
\end{center}
\end{figure}

The Dirac Yukawas influence all $\beta$-functions of the SM, in
particular the one of the top quark Yukawa coupling $y_t$. Its
$\beta$-function approximately
reads~\cite{Machacek:1983tz,*Machacek:1983fi,*Machacek:1984zw,*Arason:1991ic,*Ford:1992mv,*Luo:2002ey}
\begin{eqnarray}\label{eq:yt}
\frac{{\rm d}y_t}{{\rm d}\ln \mu} \equiv \beta_{y_t} \simeq
\frac{11}{3} y^3_t + {\rm tr} \left(Y^\dagger_\nu Y_\nu\right) y_t -
\left( \frac{17}{20} g^2_1 + \frac{9}{4} g^2_2 + 8 g^2_3 \right) y_t
\; ,
\end{eqnarray}
at energy scales above the right-handed neutrino threshold. Recall
that, in the SM, the top quark Yukawa coupling tends to be smaller
at higher energy scale due to the negative contribution of gauge
couplings in $\beta_{y_t}$ (see e.g.,~\cite{Xing:2007fb}). Since the
neutrino Yukawas contribute with a positive sign, the running
behavior of $y_t$ may be significantly modified. The RG evolution of
$y_t$ is depicted in Fig.~\ref{fig:fig_top} for some examples. An
interesting observation is that the exact value $y_t=1$ (which is
impossible in the SM) could be obtained at higher energy scales,
indicating the restoration of certain kinds of Yukawa unifications
or flavor symmetries. We see that this value can be reached for the
case of low scale seesaw with non-trivial flavor structure as well
as for the canonical case with $M_0$ being of order $10^{14}~{\rm
GeV}$.

\section{Conclusions} \label{sec:conclusion}

We have considered here the influence of neutrino Yukawa couplings
in the type I seesaw mechanism on the Higgs self-coupling. The
impact is similar to the one of the up-quarks, and thus can modify
the vacuum stability bounds at low Higgs masses. While naively this
effect decreases with decreasing seesaw scale, we have noted here
that this is not necessarily the case when the flavor structure of
the matrices is taken into account. Providing a general analysis
with a parametrization of the Yukawas, as well as giving two model
examples, we showed their impact. The Higgs mass window can become
narrower and the cutoff scale could become lower than in the
Standard Model case. It is worth to remark that the instability
bound should not be viewed as a severe problem of the model since
the SM is likely to be embedded in a more fundamental framework, and
the actual fate of the electroweak vacuum depends on the
cosmological history. In some other extensions of the SM, the
interactions between new particles and the SM Higgs coublet may lead
to a positive contribution to the Higgs quartic coupling, which
could
 stabilize the Higgs potential (see, e.g. discussions in
Ref.~\cite{EliasMiro:2012ay}). Nevertheless, whenever the seesaw
parameters are such that production of right-handed neutrinos at
colliders is possible, the effects discussed in this work can be
expected. In addition, the neutrino Yukawas may have an effect also
on the evolution of other Standard Model parameters, which we
illustrated with the example of the top quark Yukawa reaching
exactly the value 1 at high scale.

In general, seesaw contributions to the Higgs self-coupling are a
particularly simple and natural example on the impact of new physics
on fundamental properties of the Standard Model.

\begin{acknowledgments}

We are grateful to Manfred Lindner and Martin Holthausen for useful
comments. This work was supported by the ERC under the Starting
Grant MANITOP.
\end{acknowledgments}

\bibliography{bib}
\bibliographystyle{apsrevM}

\end{document}